\documentclass[epjH]{svjour}

\usepackage{graphicx}

\begin{document}

\title{Indian Amateur Astronomer R. G. Chandra:}
\subtitle{A unique AAVSO member}

\author{Sudhindra Nath Biswas\inst{1}
\and Saibal Ray\inst{2}\fnmsep\thanks{\email{saibal@iucaa.ernet.in}} \and Utpal
Mukhopadhyay\inst{3}\fnmsep\thanks{\email{utpalsbv@gmail.com}}}

\institute{Mahatma Aswini Kumar Dutta Road, Nabapally, Barasat,
North 24 Parganas, Kolkata 700126, West Bengal, India \and
Department of Physics, Government College of Engineering \& Ceramic Technology, Kolkata
700010, West Bengal, India \and Satyabharati Vidyapith, Nabapally, Barasat, North 24 Parganas,
Kolkata 700126, West Bengal, India}

\abstract{In this Note, which is a sequel on the Indian amateur astronomer 
R.G. Chandra \cite{Biswas2011a,Biswas2011b},
 we have presented some documents to reveal his three-decades relationship with 
the American Association for Variable Stars Observers and other connection to 
American astronomical societies. We have given a short account of his observations 
on different category, viz. (i) Variable stars, (ii) Nova, (iii) Meteors and 
(iv) Andromedae. His responsibilities and recognitions are also discussed.}

\maketitle

\section{Introduction}
Radha Gobinda Chandra (henceforth Chandra), a petty clerk of
Treasury Office of Jessore, received international recognition as
an amateur astronomer through his sheer love for astronomy and
meticulous observations. A detailed account of his life and works
can be found in the study of Biswas et al. \cite{Biswas2011a}.
Chandra was a member of the American Association of Variable Star
Observers (AAVSO) and regularly communicated his observational
findings to AAVSO. The main motivation of the present work is to
highlight the role of Chandra as a responsible member of AAVSO as
well as reveal some new findings which were not covered in the
aforementioned paper.

From the Eastern India, Chandra was the lone observer and member
of the AAVSO, to report his observational data, during the period
of early 20th century. He was an avid observer of the variable
stars along with the cosmic events like the eclipses, cometary
apparitions, meteor falls etc. His excellent estimations on the
brightness of variable stars were of immense importance to the
contemporary professionals of the West, in particular of the
America, for the study of some physical characteristics of the
stars.

\begin{figure}[htbp]
\centering
        \includegraphics[scale=.5]{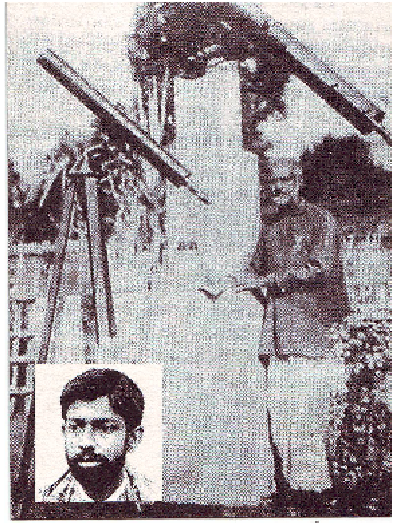}
        \caption{Photograph of aged R.G. Chandra working with his $3$ and $6$ 
inches refracting telescopes along with a photograph of young Chandra (inset)
[Reproduced from Ref. \cite{Biswas2011a}]}
   \label{Fig. 1}
\end{figure}

\section{Early Attachment with AAVSO} It is now a well known fact that  
due to his ignorance of technicalities, Chandra missed the glory
of being the discoverer of Nova Aquilae 1918 \cite{Biswas2011a}. 
The report sent by Chandra about his observation of Nova Aquilae reached 
Charles Pickering (1846-1919) six months after its first reporting.
Although, by the time, significance of the report had faded out
from the professional astronomers' point of view, yet the said
report drew considerable attention of Pickering who promptly
realized the potentiality of Chandra as a good observer.
Accordingly, on 14 November 1918 he dispatched an inspiring letter
accompanied by a few valuable books, star map, Revised Harvard
Photometry and some literatures on the Nova Aquilae. He also made
necessary arrangement for enlisting Chandra as a member of the
AAVSO, in recognition of his ability equal to the standard of
international observers. But it should be mentioned here that even
before Chandra sent his `report' on the `new star' (i.e. Nova
Aquilae) to Pickering, he made communication with the AAVSO by
expressing his intension to take part in the process of estimating
variables. This point becomes evident from a part of the letter
dated 8 July 1918, written by S.I. Belly \cite{Campbell1932a}, on behalf of
the AAVSO: 

{\it ``Situated as you are six hours East of Greenwich,
your station should prove of great help in the observation of
these variables ...}".

Since 1919 Chandra could contribute his data collected from the
observation of variable stars to the journal of several learned
societies of Europe and America. In connection to the AAVSO
history Part-4: `The AAVSO and International Cooperation' it has
been reported in the AAVSO portal, for the members between 1911
and 1921, that ``{\it Other early international observers were:
Radha G. Chandra of Bagchar, India, who made over 49,700
observations ...}''. So, probably Chandra became the member of
AAVSO around the year 1919.

\section{AAVSO Telescope} Due to his advantageous location of observation
and quality of estimation of magnitude of variables, Chandra's reports
earned immense importance from the professionals for their studies
on the physical characteristics of stars. In order to receive
better reports on fainter variables, the authorities of AAVSO were
contemplating to lend him a more powerful telescope. Accordingly
Leon Campbell, the Chairman of its Telescope Committee, informed
Chandra about their intension of lending him a $6\frac{1}{4}$-inch
refractor through the following letter:

\vspace{1.0cm}

{\it HARVARD COLLEGE OBSERVATORY CAMBRIDGE, MAS\\

\vspace{.50cm}

Mr.R.G.Chandra~~~~~~~~~~~~~~~~~~~~~~~~~~~~August 12, 1924.\\
~~~Bagchar,\\
~~~Jessore P.O. India\\

My Dear Mr. Chandra,

For a long time I have been very desirous of securing for you the
use of a larger telescope than you have. At last this seems about
to be realized. Our patron and friend Mr. C. W. Elmer of N.Y. has
just turned over to the Association his $6\frac{1}{4}$ lens in
tube with finder and ocular and cradle clamps for attaching the
tube to a mounting. The lens is very good one and should enable
one to see much fainter stars than with a three inch instrument.

Now if you can see your way clear to provide some sort of
mounting, either temporary or permanent, the telescope committee
is willing to let you have the loan of this splendid equipment as
described above.

... Just how to best arrange for such a loan is the serious
question. As long as you live and will keep the instrument in
reasonable use for AAVSO, variable star observing, the
telescope can be considered as virtually yours. The difficulty
comes in the case of your death. What assure can be had that the
equipment would be restored to the Association, either to some
other observer in Asia, Europe or U.S.A.? This is what bothers us.
The equipment is valued at 500 dollars and as long as you keep it
in good use, we shall feel well repaid for our efforts in lending
to you.

I might suggest that you own three inch would make a very
desirable additional finder, especially if you have no circles at
first. The Association could defray the initial cost of
transportation, asking you to repay the Association as you could.

If you decide that the loan of this equipment is practical, that
you can provide some sort of mounting for the present at least and
will agree to use it exclusively for AAVSO observing, will take
good care of it, and reimburse the Association later for the
transportation expenses and provide for its return to the
Association or it authorized agent upon your demise or inability
to make further use of it, let me know and I'll start steps for
having it sent to you at once.\\

With best wishes and kindest regards, I am,\\

Faithfully yours\\

Leon Campbell\\ Chairman Telescope Committee.}

\vspace{.50cm}

As the mounting, together with the said telescope reached Bagchar
in damaged condition in the year 1926, Chandra had to repair the
mounting before he began to use the $6\frac{1}{4}$-inch from 1928.

\section{Chandra as an AAVSO-Observer}
\subsection{Observation of Variable Stars} The data of instant
brightness collected from the systematic observations of variable
stars enable astronomers to reveal several physical
characteristics, viz., the mass, radius, temperature, luminosity,
composition, both the internal and external structures and
evolution of the stars. In order to achieve a more reliable
result, these data are required to be collected in sufficient
quantity from various stations scattered at far and wide distances
all over the globe. In this regard, there were good numbers of
variable star observers contemporary to Chandra, from Europe and
America on the Western longitudes, but not sufficient on the
Eastern side to report their estimations on variables to the
AAVSO. From India Chandra \cite{Campbell1925} of Jessore, Bengal ($23^0~10'N$,
$89^0~10'E$), George E. Jones \cite{Chakrabarty1999} of Mussoorie, U.P. ($30^0~27'N$,
$78^0~06'E$) and M. K. Bappu \cite{Campbell1926} of Begampeth, Hyderabad
($17^0~10'N$, $70^0~20'E$) were the only three to contribute `estimates'
to the said Association, since 1919, 1924 and around 1927
respectively.

Leon Campbell (1881-1951) was elected President of AAVSO in 1919,
the year from which Chandra began to report his estimations on
variables to the same at regular intervals. Thus in the process of
serving the `Association' by these two astronomers in different
roles from the opposite hemispheres of the Earth for years
together, both Campbell and Chandra were in close contact until
the former died in 1951.

The quality of work done by Chandra can be realized from the
remark made in a letter \cite{Campbell1926}, dated 20 June 1922, addressed to
him by Harlow Shapley (1885-1972), Director of the Harvard
College Observatory:\\

{\it ``May I add a personal word of congratulation for the good
work you have been doing in the observation of long period
variable stars. Your longitude is of considerable importance in
this work.}"\\

The available data reported by Chandra to the AAVSO and
subsequently published in its Monthly Reports, reveal how much
laborious and ardent he was as an observer of the `Association'.
For instance, during Campbell's tenure as Recording Secretary of
AAVSO, Chandra's month-wise reporting of estimates of the
brightness of a total of 1685 variable stars was the largest among
such number in any single year ending October 1926. Though
analysis of the said `Report' of the actual number of estimations
recorded by observing Julian day-wise shows that Chandra made 1682
estimations on about 160 variables, during the year ending
September 1926 \cite{Campbell1926} [see Table-1]. Although Chandra
himself made highest number of $247$ estimations in the month of
January 1926, yet he was lauded later as one of the three major
contributors of reports on variables to the AAVSO. As such
Campbell as the Recording Secretary, while publishing those
reports sent by $25$ observers who made $1358$ estimations on $313$
stars for the month of April 1926, in the Monthly Report
of the AAVSO \cite{Campbell1932a} made the following comments on 12 May 1926:\\

{\it ``Most commendable list have been received the past month
from Messrs Peltier, Chandra and Waterfield, their combined
records totaling more than half the number of all the observations
combined therein.}''

\begin{table}
\caption{The highest number of estimates made by Chandra on the
variables in any single year ending October 1926}\centering
\bigskip
{\small
\begin{tabular}{@{}llrrrrlrlr@{}}
\hline \\[-9pt] Months & Julian Day & Observed no. of Stars  & No. of actual estimation & No. of obs. reported\\ \hline \\ [-6pt]

1925 Oct        & 2424 424-455      &  67  &  118 & -\\
~~~~~~~ Nov     & ~~~~~~~456-485     &  109 &  193 & 105\\
~~~~~~~ Dec     & ~~~~~~~486-516     &  108 &  169 & 223\\
1926 Jan        & ~~~~~~~517-547     &  115 &  247 & 178\\
~~~~~~~ Feb     & ~~~~~~~548-575     &  109 &  220 & 162\\
~~~~~~~ Mar     & ~~~~~~~576-606     &  76  &  185 & 226\\
~~~~~~~ Apr     & ~~~~~~~607-636     &  87  &  169 & 175\\
~~~~~~~ May     & ~~~~~~~637-667     &  49  &  81  & -\\
~~~~~~~ Jun     & ~~~~~~~668-697     &  53  &  77  & 312\\
~~~~~~~ Jul     & ~~~~~~~698-728     &  52  &  74  & -\\
~~~~~~~ Aug     & ~~~~~~~729-759     &  59  &  87  & 139\\
~~~~~~~ Sep     & ~~~~~~~760-789     &  48  &  62  & 83\\
1926 Oct        & ~~~~~~~790-820      &  -   &  -   & 82\\
\hline
    TOTAL       &                   &      &  1982 & 1685\\
\hline

\end{tabular}
}
\end{table}

In the month of April 1926, L.C. Peltier, R.G. Chandra and W.F.
H. Waterfield reported $259$, $175$ and $237$ estimations respectively.
Actually, total of these estimations are not `more than half' but
`very nearly equal to the half' of the number 1358\footnote{Interested 
readers may consult the ANNEXURE~I \cite{Bandyopadhyay1991} for some of the 
contributions on Variable Stars by Chandra in Annual and Monthly Reports of AAVSO (1920-21)}. 

Thus Chandra had been reporting his estimates on variables in the journals of
AAVSO, BAA, de L'Observatoire de Lyon etc. and gradually his
excellence of work was also waxing with confidence. Within a few
years he became a competent and indispensable member in the
observational astronomy. In course of time, he was contemplating
to observe fainter stars of higher magnitudes than those were
capable to view through his $3$-inch refractor. At the same time,
he came in contact with J. H. Logan of Dallas, Texas who began to
report his `estimates' probably from October 1926 to the AAVSO as
its member and was in possession of an 11-inch telescope for his
observation \cite{Campbell1928}. In reply to some quarries of Chandra which
include the size of telescope required for viewing the stars of
higher magnitudes, Logan sent a long correspondence \cite{Hoffleit1942}. From
this correspondence, the answer related to information of the size
of telescope, is given below:\\

{\it ``You asked me about the size of telescope necessary to see
stars to the $16^{th}$ or $17^{th}$ magnitudes. There is a formula which can
determine this:

\begin{equation}
Log~aperture~in~inches \times 5 + 9.2''
\end{equation}}

The formula (1) implies that the faintest star can be viewed:\\
(a) through his $3$-inch is of $Log~3 \times 5 + 9.2$ or $11.59$
mag, and (b) through the $6\frac{1}{4}$-inch lent to Chandra by
the AAVSO, is of $Log~6.25 \times 5 + 9.2$ or $13.18$ mag. Due to
such variations in the capabilities of different sizes of
instruments Chandra could not view the faint stars beyond the
$12.6$ magnitude through his 3-inch refractor. But with the
lending of the $6\frac{1}{4}$-inch refractor from the AAVSO he
could now detect faint stars up to $14.1$ magnitudes [see
Table-2].

\begin{table}
\caption{The magnitude of some of the most faint stars Chandra
observed: (A) Prior to and (B) Post acquisition of
$6\frac{1}{4}$-inch telescope}\centering
\bigskip
{\small
\begin{tabular}{@{}llrrrrlrlr@{}}
\hline \\[-9pt]
& Observed & Stars & Date of observation & Report to AAVSO\\
& magnitude&       & Julian Day Gregorian Day& Year Page\\ \hline \\
[-6pt]
A & 12.6 & 185032 RX LYRAE & 242 2984~~~~ 21 Oct 1921 & 1922~~~ 20\\
  & 12.5 & 190925 S LYRAE  & ~~~~~~3613~~~~ 12 Jul 1923 & 1923~~~ 98\\
  & 12.4 & 050492 U ORIONIS& ~~~~~~4262~~~~ 21 Apr 1925 & 1925~~~ 93\\
  & 12.3 & 163264 R DRACONIS& ~~~~~~3315~~~~ 17 Sep 1923 & 1923~~~ 07\\
  & 12.1 & 180531 T HERCULIS& ~~~~~~2960~~~~ 27 Sep 1921 & 1922~~~ 08\\
  \hline \\
B & 14.1 & 074922 U GEMINORUM& ~~~~~~5649~~~~ 06 Feb 1929 & 1929~~~ 54\\
  & 14.0 & 042209 R TAURI& ~~~~~~6712~~~~ 05 Jan 1932 & 1932~~~ 50\\
  & 14.0 & 060547 SS AURIGAE& ~~~~~~5622~~~~ 10 Jan 1929 & 1929~~~ 44\\
  & 13.7 & 190967 U DRACONIS& ~~~~~~5448~~~~ 20 Jul 1928 & 1928~~~ 81\\
  & 13.5 & 004746a RV CASSIOPEIAE& ~~~~~~6635~~~~ 20 Oct 1931 & 1932~~~ 02\\
\hline

\end{tabular}
}
\end{table}

The AAVSO authorities, in particular, Leon Campbell never missed
any opportunity to accord appreciation for any worthy work of
Chandra for the advancement of observational astronomy. In one
such occasion, while reporting contributions of members of the
AAVSO for the year ending October 1932, Campbell \cite{Campbell1932b} 
highlighted the work of a few including Chandra out of $80$ observers from various
parts of the globe: \\

``{\it Of the $33000$ observations, $21000$ were
contributed by $10$ observers. Of these ten, Peltier heads the list
with slightly over $4,000$ observations, Jones is a close second,
and Lacchini comes third. Next in order are Ahnert of Germany,
Chandra of India, Baldwin of Australia, and Ensor of South
Africa}."\\

Actually, Chandra reported $1655$ estimations, including the highest
number of $487$ estimations in a single month of March, during that
period.

Due to World War II, during the periods September 1939 - September 1945, 
many astronomers were on war duty and could not continue their observations. 
At such crucial time Chandra played an important role for the cause
of observational astronomy. How nicely he discharged his
responsibilities, can be revealed from the following `News Notes'
published by Dorrit Hoffleit \cite{Hoffleit1942}, in the Sky and Telescope
under the title `Variable Star
Observations from India':\\

{\it ``The Recorder of the American Association of Variable Star Observers, Leon Campbell,
reports that the observations of variable stars have continued to
flow in from India, despite the difficulties and delay in
transportation, and despite the great menace of the war. He has
received large number of estimates of the brightness of variable
stars from R.G. Chandra, of Bagchar, about every six months ...~.}"\\

We would like to hear on this particular issue of war-affected period from 
Campbell himself \cite{Campbell1946}:\\

{\it ``R.G. Chandra of Bagchar, India, has been an observer since 1920, 
with more than 30,000 observations to his credit. Here again, the war 
greatly interfered with his work during the past five years.''}\\

Some time prior to August 1945, M. K. Bappu was in need of the
service of a telescope. As a genuine astronomer Chandra, at the
request of Campbell, gladly agreed to lend his $3$-inch telescope
to Bappu. For this generous act Bappu conveyed his thanks to
Chandra through the following letter:

\vspace{1.0cm}

{\it Begampet~~~~~~~~~~~~~~~~~~~~~~~~~~~~~~~~~~~~14th August, '45\\

Dear Mr. Chandra,

It is very kind of you to offer me the loan of your $3"$ Refractor
with its accessories so as to enable me to continue my
observations of variable stars and I thank you heartily for the
same. I am also grateful to Prof. Campbell for kindly recommending
me to you. But I am sorry I cannot avail your generous offer
immediately. As I intend going to Malabar during the first week of
September and do not expect to be back before December, I do not
think it will serve any purpose if I request you to send it to me
now only. As soon as I return I will write to you.

Of course will meet all the charges required for its dispatch, and
after it reaches me safe here I shall hold myself responsible for
maintaining its condition and its safe return to you in time. The
risk of loss or damage in transit, I believe even in this time,
will be bored by the Railway, if article is properly packed an
adequately insured. However I hope by the time I return and
request you to send me the telescope, normal condition will
prevail and the danger of the transit damage in transport will
disappear.

In the mean time I will learn for how much you will insure the
parcel so that I may have an idea what custom duty I shall have to
pay on the article at this end. This information will also help me
to correspond with the authorities for an exemption from custom
duty as the instrument is coming only on loan and not going to
remain here permanently. The Govt. charge $5$ p.c. custom duty on
all articles imported.

Once again thanking you heartily,\\

I remain\\

Sincerely Yours,\\

M. K. Bapuu}

\vspace{0.50cm}

Also, Campbell expressed his thanks to Chandra for complying with his
request through the letter given below:

\vspace{1.0cm}

{\it HARVARD COLLEGE OBSERVATORY CAMBRIDGE, MASS

\vspace{0.50cm}

Mr. R.G.Chandra ~~~~~~~~~~~~~~~~~~~~~~~~~~~~February 2, 1946\\
Bagchar, Jessore P.O., India.
\\

Dear Mr. Chandra,

It is certainly generous of you to place on loan to Mr. Bappu the
three inch telescope, and I thank you on behalf of the
Association, as well as on my behalf. Mr. Bappu was an excellent
observer when he had access to a large telescope, and I am looking
forward to future. You might be interested to know that his son is
also very much interested in variable star observing, and to date
he has been contributing observations made with the naked eye. Now
things are getting back to normal, I am anticipating the receipt
of more observations from your section of the country, especially
so in view of the fact that we are approaching the time when
one-millionth observation, as made by AAVSO, will be contributed
before June first. Perhaps you may be proved to be the one who
makes that one-millionth observation. Time only can tell. With
kindest personal regards, and best wishes for your continued
success, and ever mindful of the long record which you have made
in the matter of observing variables for the AAVSO,\\

I am,\\

Very sincerely yours,\\

Leon Campbell\\
Recorder, AAVSO.}

\vspace{0.50cm}

In the same letter Campbell wished Chandra to be able for making
the one-millionth observation on a variable star. However, in the
long run the one-millionth observation was achieved not by Chandra
but Jocelyn R. Gill in April 1946 \cite{Hoffleit1946}.

\subsection{Observation of Nova} As an assiduous
observer Chandra had observed a few novae besides his detection of
the $Nova~Aquilae~1918$ or $V~603~Aquilae$ as its first observer. His
ignorance about the formalities related to discovery of a
celestial object deprived him from the glory of Nova discovery.
The incident inspired him to continue for searching of a nova.
When Campbell came to know about his effort, he wrote Chandra some
encouraging words in a letter on 16 June 1921: \\

{\it ``You have taken up nova search in a good sprit and I hope you may be
rewarded some day with a real Nova discovery.}" \\

Along with the usual observation on variables he also estimated the brightness of
$184300~Nova~Aquilae$ in the year 1918. It has been found in the {\it
Monthly Report} of the AAVSO for the years 1921, 1922, 1925 and
1926 that during the period of two years from September 1921,
Chandra made 15 estimations (Table-3) \cite{Eaton1922} and that of from
September 1924, he made 19 estimations on the brightness of the
Nova. From these it is found that the brightness of the Nova,
estimated during the period of two years from September 1921, were
slightly fluctuating in between $9.00$ mag and $10.3$ mag, whereas
those for the next succeeding years remained almost steady at
about $10.4$ mag which is close to its normal brightness of $10.5$
mag. Perhaps Chandra was able to observe the transition stage of
the Nova, before it declined to its minimum brightness at steady
state.

\begin{table}
\caption{The brightness of the Nova V 603 Aquilae estimated by
Chandra during the period from September 1921 to September
1923}\centering
\bigskip
{\small
\begin{tabular}{@{}llrrrrlrlr@{}}
\hline \\[-9pt] Julian Day & Gregorian Day  & Magnitude\\ \hline \\ [-6pt]

              242 2940.2  &  1921, Sep 07 & 9.0\\

              ~~~~~~2984.0  &  ~~~~~~~Oct  21 & 9.3\\

               ~~~~~~3016.0  &  ~~~~~~~Nov 22  & 9.2\\

               ~~~~~~3142.5  &  1922, Mar 28  & 9.4\\

               ~~~~~~3178.4  &  ~~~~~~~May 03  & 9.9\\

               ~~~~~~3196.2  &  ~~~~~~~May 21  & 10.0\\

               ~~~~~~3238.2  &  ~~~~~~~Jul 02  & 10.0\\

               ~~~~~~3292.2  &  ~~~~~~~Aug 25  & 10.0\\

               ~~~~~~3312.2  &  ~~~~~~~Sep 14 & 9.9\\

               ~~~~~~3339.1  &  ~~~~~~~Oct 11  & 9.7\\

               ~~~~~~3341.1  &  ~~~~~~~Oct 13  & 9.6\\

               ~~~~~~3563.5  &  1923, May 23   & 10.2\\

               ~~~~~~3608.1  &  ~~~~~~~Jul 07  & 10.2\\

               ~~~~~~3643.1  &  ~~~~~~~Aug 11  & 10.3\\

               ~~~~~~3677.2  &  ~~~~~~~Sep 14 & 10.3\\

\hline
\end{tabular}
}
\end{table}

Another famous nova of the 20th century, $Nova~Hercules~1934$ or $DQ~Hercules$ \cite{Campbell1925} 
was discovered by a British amateur astronomer
Prentice on 13 December 1934 \cite{Struve1962}. However, from the text of
Campbell's letter, it is apparent that Chandra observed the nova
DQ Hercules to oscillate in its postnova stage and the same
phenomena continued for a long time. The letter, dated 29 March
1935, written by Campbell is as follows: \\

``{\it I am pleased to acknowledge your postal card of the 16th Jny., 
concerning the first estimates of Nova Hercules. I am glad to note that 
you have kept watch on this star and have secured such a continuous series
of observations. The fluctuations noted in the star certainly been
real and these have not ceased even up to present time.}"\\

We also notice about Chandra's observation from another source \cite{Mayall1963} report:\\

{\it ``In the early morning hours of December 13, 1934, a nova of magnitude $3.3$ 
was discovered visually by J.P.M. Prentice, an English amateur. 
It was also discovered independently by three members of the A.A.V.S.O.: 
Leslie Peltier, who was travelling in Arkansas; R.G. Chandra in Bagchar, 
India; and Margaret Harwood, in Nantucket, Massachusetts''.}

\subsection{Observation of Meteors} That Chandra also was a
competent meteor observer can be revealed from the response to one
of his report on such observation sent to the American Meteor
Society, Lede Mccormick Observatory, University of Virginia. In
reply, from the Headquarters of the `Society', dated
5 February 1926, not only his excellent observations were
appreciated but also he was invited to join the said Society, in
the following words:\\

{\it ``Your note enclosing to excellent observations of telescope
meteors reached me today .... As you are evidently a real worker
in astronomy, how would you like to join the Am. Meteor Society?''}\\

In a letter, dated 23 June 1927, Willard J. Fisher \cite{Chandra1985} of
the Harvard College Observatory (H.C.O.) Cambridge, Massachusetts 
wanted to know from Chandra about the recorded knowledge
of ancient Indians on the Leonid meteors. From a paper published
by Hubert A. Newton \cite{Newton1865}, Fisher found such knowledge from
the countries like China, Arab and Europe, except India. He was
sure about the fact that such wonderful celestial event cannot
evade the knowledge of ancient Indians. In spite of his best
effort, Chandra  was sorry for not being able to not reply the
quarries of his request. Yet, Chandra sent him about the incidents
of meteor falls and cometery apparitions available in the Indian
epics, Mahabharata and Ramayana. In the case of another letter by Chandra of 
dated 7 September 1927, containing some reports on the
Indian records of shooting star shower, Fisher gladly replied on 3
October 1927 as follows:\\

{\it ``Dear Mr. Chandra, 

I have yours about Indian records of shooting
star shower, dated Sept. 7, for which I am greatly obliged. I must
confess to a very complete ignorance of the astronomical
literature to which you refer, and I learn for the first time
about the ruling motions for the study of astronomy in ancient in
India.

I am certainly pleased that you are willing to take further
trouble with regard to these matters. Could you not work up the
result of your search, negative results as well as positive, into
a paper? Such a paper would be welcome, I am sure, and I have no
doubt that a place for its publication could be found, in India,
or failing that here in this country.\\

Yours very truly,

Willard Fisher."}\\

From available records, it is found that Chandra reported on his
successful observation of $24$ meteors during 1928-30 and another 9
in 1937. Also, the following report on his observations of 22
meteors was published by Charles P. Oliver in 1931 \cite{Oliver1931}: \\

{\it ``We have also received, too late for last years' report, observation
of 17 telescopic and 5 casual meteors by R. G. Chandra of Jessore,
India."}

\subsection{Observation of Andromedae} 
It is a surprizing information that Chandra even observed Andromeda 
by using his $3$ and $6$ inches refracting telescopes. The record is 
in the paper written by J. van der Bilt. It was published in the very 
prestigious and authentic British journal `Monthly Notices of the 
Royal Astronomical Society' with the title `The light-variation of $V22=SV Andromedae$' 
\cite{Bilt1934}. In the Introductory part of the paper it has been mentioned by Bilt that -\\

{\it ``From the list of A.A.V.S.O. observers I have omitted the names of those 
who have contributed only a single observation. As to the others I consider it a 
matter of courtesy not to treat them anonymously, but to give a list of their names 
and particulars about their instruments; especially since it appears that the 
mean deviation of the estimates of these observers (taken as a group) from 
the final light-curve does not differ appreciably from that of my own estimates''.}\\

\begin{table}[htbp]
        \includegraphics[scale=.60]{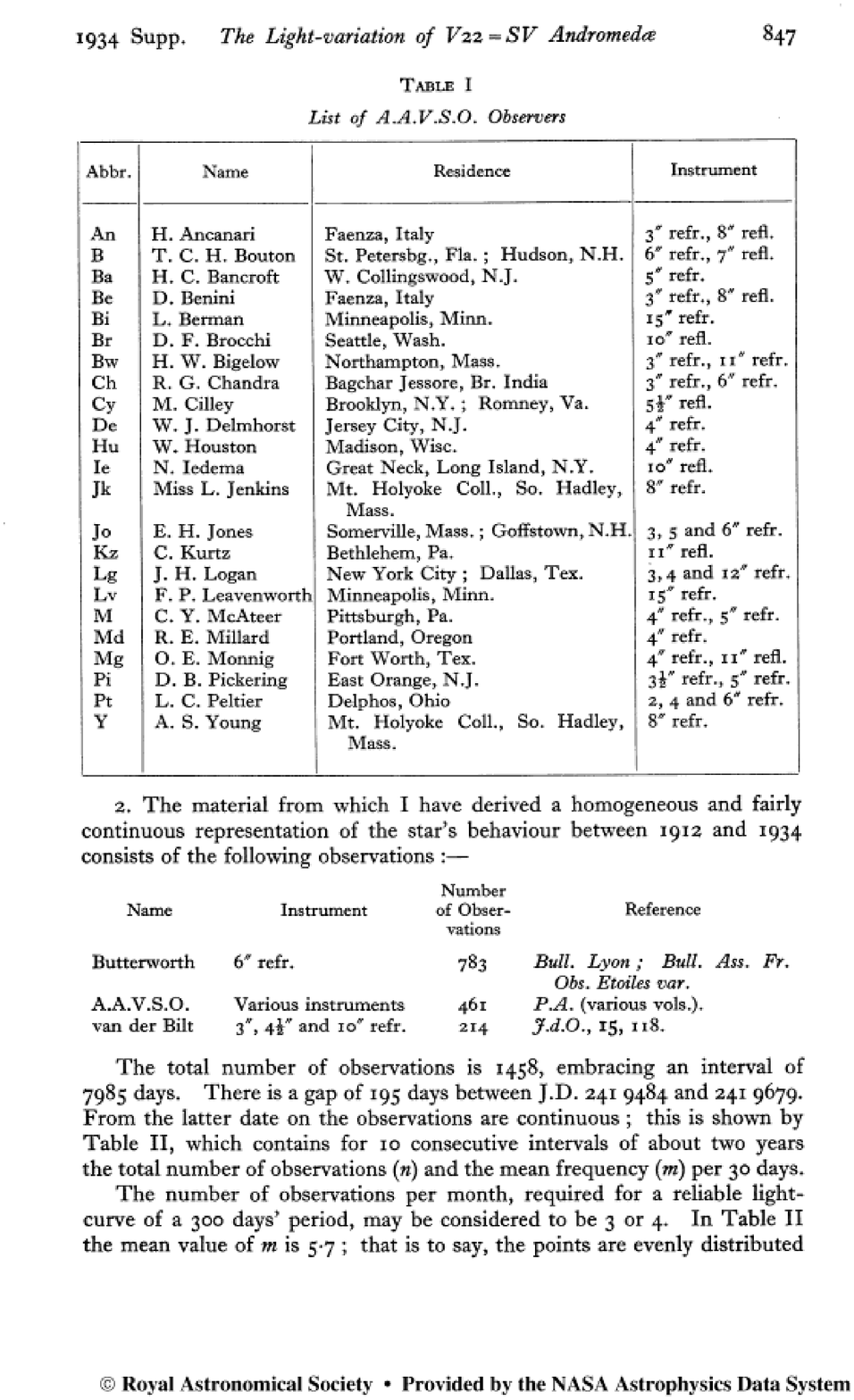}
        \caption{Photocopy of the list of AAVSO observers on Andromedae}
   \label{Table 4}
\end{table}

From this statement about the list of the AAVSO-observers (see Table 4) 
it is clear that Chandra other than a single observation contributed lots of data 
on Andromedae. Also it is obvious from Bilt's remarks that Chandra's observational 
data and hence the estimates on mean deviation from the final light-curve were 
comparable to the other World-class observers including the author of the paper himself.

\section{A Responsible Person} 
We see a Variable Star Notes from the AAVSO by Campbell \cite{Campbell1936}
as follows:\\

{\it ``... we are glad to welcome back to the ranks of active observers 
Mr. R.G. Chandra, of Bagchar, India, who, because of ill health, 
has been forced to take a vacation of several months from observing.''}\\

So, as Chandra became incapacitated to
perform any kind of intellectual activities, he on the advice of
AAVSO authorities, handed over the $6\frac{1}{4}$-inch refractor
to M. K. Vainu Bappu (1927-82), in 1958. The said telescope has
been installed at the entrance of Vainu Bappu Observatory at
Kavalur ($78^0~49.6'E$,~$12^0~34.6'N$, Altitude - $725$~m) in the North Arcot
district of Tamil Nadu India, under the direct supervision of the
IIA. This handing over of the telescope shows Chandra's sense of
responsibility towards astronomy. This dutifulness prompted him
not to keep the equipment in his own custody when he was unable to
make proper use of it.

Apart from observation of celestial objects, Chandra was a keen
reader of astronomical journals, including the Sky and Telescope.
After reading its March, 1951, issue, he became curious about the
origin of naming the days in a week. In order to know about the
`origin of naming', he wrote the following letter to the Editor of
that journal and it was published in its January 1952, issue:\\

{\it ``Sir,

In terminology Talks, March, 1951, under `The Week', I read `we
readily recognize the planetary origin of the names ...

But was there any reason, cause, or rule to arrange the names of
the days in alternate order of one outer and one inner planet? For
example, after Moon, Dies Lunie, the outer Planet Marsis taken for
Dies Martis, the inner planet Mercury is for Dies Mercurii, then
the outer planet Jupiter for Dies Jovis. Venus was taken for Dies
Veneris and next Saturn as Dies Saturni. The Moon is the inner
bright orb at new Moon and the outer on at full Moon, therefore I
assume Dies Lunae after Solis. But for the others? I shall be glad
if anybody could find out the cause.\\

R.G.CHANDRA\\

Sarkarbati,\\
P.O. Sukchar,\\
Dist. 24 Pargana,\\
India (West Bengal).}

\vspace{0.5cm}

In response, Chandra received a brief reply in March 1952, issue
from Edgar W.Woolard of Washington D.C. given below:\\

{\it ``The names of the seven planets of the ancient were attached to
the 24 hours of the day as ruler of these hours, in order of
descending distance from the Earth in the ancient Greek geocentric
system; the planet that rule the first hour was the ruler of the
day and the day was named after this planet."}\\

M.W. Mayall made report in the {\it Popular Astronomy} \cite{Mayall1951} 
about the contribution by Chandra with his ill health even in 1951. 
We note from the Table 5 that he made $73$ observations with $93$ estimates.

\begin{table}[htbp]
        \includegraphics[scale=.60]{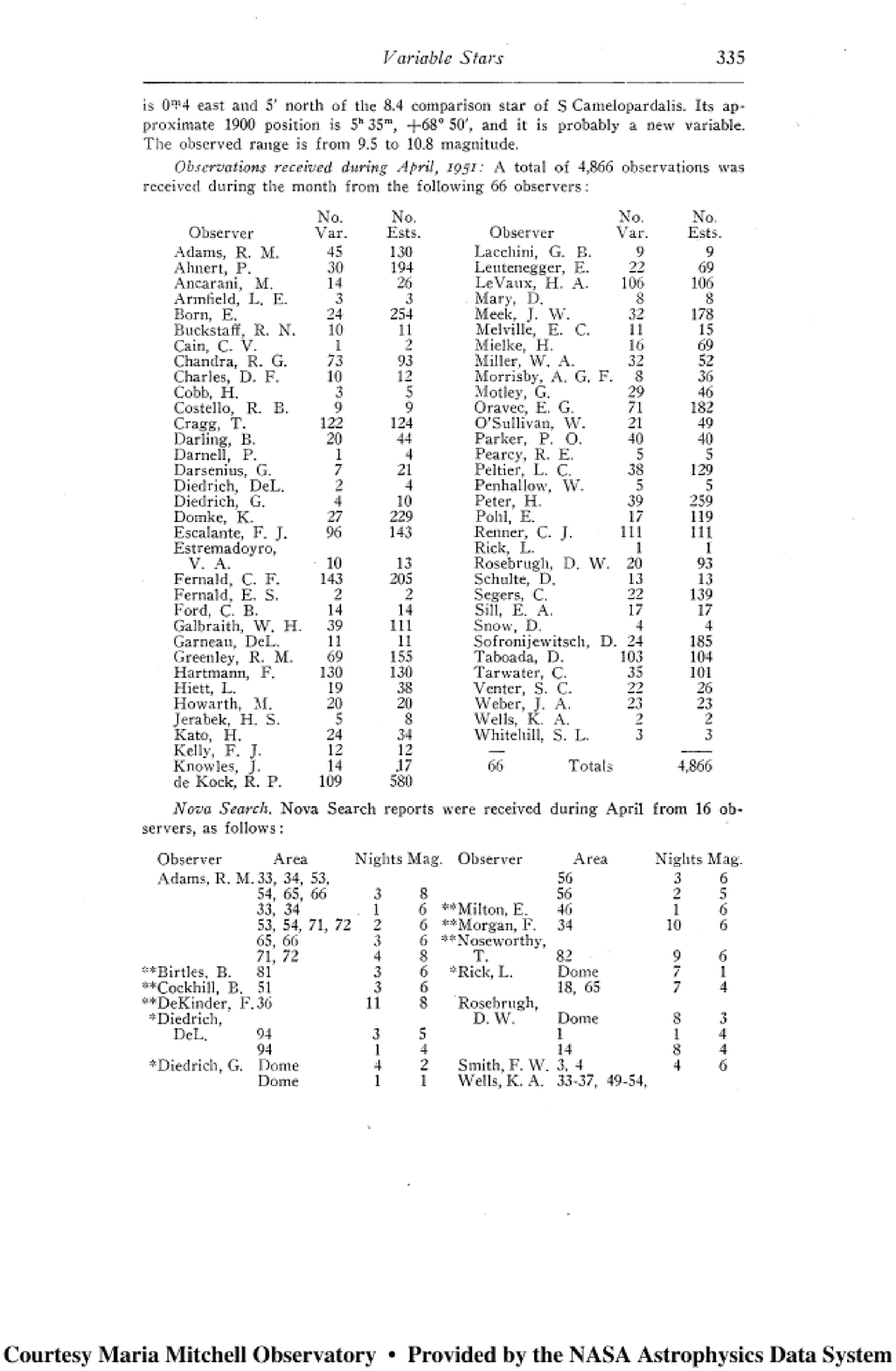}
        \caption{Photocopy of the list of observations on Variable Stars 
during April 1951 by Chandra}
   \label{Table 5}
\end{table}

\section{Recognition from AAVSO} 
Apart from tendering appreciation from time to time for his 
outstanding work in estimations of variables, Chandra was 
accorded some rare distinction by the AAVSO. In connection to 
this we would like to quote 
here from the {\it Centenial
History of the American Association of Variable Star Observers},
\cite{Williams2011}:\\

{\it ``The case of Radha G. Chandra of Bagchar, Jessore, India, 
illustrates a growing international influence in this period. 
A clerk in the tax collector's office by day, he enjoyed watching 
the stars at night. When Chandra made an independent discovery of 
Nova Aquila 1918 and reported his results to HCO, Campbell recommended 
him as a member of the AAVSO.''}\\

On one such occasion, 
in 1932, Campbell intimated him that he was relieved from paying 
annual subscription to the `Association' in the following words \cite{Campbell1941}:\\
   
{\it ``In view of the splendid record which you . . .  have made 
during the past years in connection with variable star observing 
and in view of the fact a good friend of the Association decided 
to defray the annual dues ...''}.\\

The most prestigious reward was his election as a honourary member
of the AAVSO at its Annual Meeting in 1947 at Harvard.
This news was reported in the {\it Sky and Telescope} \cite{Hoffleit1947} 
in the following manner: \\

{\it ``Two outstanding observers were elected to honorary
membership: Rev. T. C. Bouton, St Petersburg, Fla, and R. G.
Chandra of Bagchar, India.}"\\

In connection to the opportunity for international involvement, 
the AAVSO Council, held in November 1920, gave extraordinary 
importance and recognition to Chandra. The available Record 
\cite{Williams2011} on this matter is like this:\\

{\it ``... the Council considered a suggestion from Indian member 
and regular contributor R.G. Chandra that the AAVSO undertake a 
statistical analysis of its obsrvations database on a regular basis 
but concluded that the financial resources available would not 
permit such an undertaking''}.\\

Another remark can be quoted here as made in a letter, dated 12 December 1950, 
by Harlow Shapley, Director of the Harvard College Observatory \cite{Chakrabarty1994}:\\

{\it ``The American Association of Variable Star
Observers, with Head-quarters at the Harvard Observatory, 
is honored to salute you as one of its important contributors from abroad.}"

\section{Conclusion} Though, the academic attainment of Chandra was
not conspicuous, yet by dint of his determination and dexterity he
achieved the distinction of becoming one of the best observer
astronomers among his contemporary amateurs in the world. In the
series of Harvard radio talks broadcast from the station WRUL,
Leon Campbell as Recorder of the AAVSO, appropriately narrated on
8 March 1941, about the amateur astronomer like Chandra and
specially mentioned his rich contribution as an observer, in the
following lines \cite{Campbell1941}: \\

{\it ``By amateur astronomer we mean one who
makes astronomy his hobby, or avocation, rather than his vocation.
He it is who observes the stars for the love and the thrill of the
thing; and he surely finds plenty of thrills, whether in
snow-covered yards in a New England city, the wide expanses of our
Middle-West, or in more salubrious climates of Italy and India. He
it is who drag out his telescope each clear evening to see ``what's
up" in variable star activities, and to note if a particular star
had faded considerably in light since the last time he looked, or
if another star suddenly increased in brightness since last night, 
or perhaps this particular amateur, variable star observer has
had erected in a shelter atop his home, or in his back yard, a
more pretentious telescope, fitted with setting circles and
electric clock-drive with which he can turn upon a pet variable
star and find the special star right there in the center of the
field of view.

... In foreign countries we have Radha G. Chandra, Customs
Official of Bagchar, India. Mr. Chandra, now in his sixtieth year,
who has been aiding in the variable star work since 1919, has
accumulated probably more observations on variable stars than any
other A.A.V.S.O. foreign observer, well over 50,000."}\\

At the fag-end of his of his career as an observer Chandra
received the following felicitation from Harlow Shapley, then
Director of the Harvard College Observatory \cite{Chakrabarty1999}:

\vspace{1.0cm}

{\it HARVARD COLLEGE OBSERVETORY\\
CAMBRIDGE  38, MASSACHUSETTS\\

``Dear Mr.Chandra,~~~~~~~~~~~~~~~~~~~~~~~~~~~~~~~~~~~~~~~~~December
12, 1950

The American Association of Variable Star Observers, with
headquarters at the Harvard Observatory, is honored to salute you
as one of its important contributors from abroad. On the occasion
of the thirty-ninth annual meeting we of the Association and of
the Harvard Observatory joined in recognizing that astronomical
work such as you accomplish is a significant contribution in the
cause of international good-will and cooperation. From seventeen
different countries come the systematic measures of the Sun and
the variable stars that are of supra-national interest to all of
us; we are in a sense, a Stella United Nations. 

We wish you continued success in you work during this fortieth year of the
Association, and hope that in this modest enterprize of ours we
are providing an example of cooperation. The members of the AAVSO
are showing how the people the world over can substitute for
strife and suspicion this intellectual and technical collaboration
and how we can build the scientific friendship that are essential
for a continued civilization.''}\\

There are some other letters\footnote{See the photocopies in the Ref. \cite{Biswas2011a}} 
written by AAVSO personnel, e.g. Campbell, Fisher etc. time to time to 
Chandra which reveal the fact that the relationship between them was 
not only official based on astronomical observations 
but also very cordial and friendly.

\section*{Acknowledgement} The authors are thankful to both Late
Kalidas Chandra, the eldest son and Sisir Kumar Chandra, the
grandson of R. G. Chandra for enlightening about
many aspects of life and activities of Chandra.
Hearty thanks to Prasenjit Basu, the Future Institute of
Engineering and Management, Kolkata, for helping in preparation
the manuscript. We all are grateful to Professor J.V. Narlikar 
for his special encouragement to write this article.

{}


\begin{thebibliography}{}

\bibitem[Biswas et al. 2011a]{Biswas2011a} Biswas, S.N., Mukhopadhyay, U. and Ray, S. 2011a. {\it Ind. J. Hist.
           Sc.}, {\bf 46}, p. 483.

\bibitem[Biswas et al. 2011b]{Biswas2011b} Biswas, S.N., Mukhopadhyay, U. and Ray, S. 2011b. {\it arXiv: 1104.1391 [physics.hist-ph]}.

\bibitem[Campbell 1932a]{Campbell1932a} Campbell, L. 1932a. {\it Monthly Reports and Annual Report of the AAVSO}, p. 130.

\bibitem[Campbell 1925]{Campbell1925} Campbell, L. 1925. {\it Monthly Report of the AAVSO}, p. 24.

\bibitem[Chakrabarty 1999]{Chakrabarty1999} Chakrabarty, R. 1999. {\it Jyotirbijnani Radhagobinda}, Puthipatra Pvt. Ltd.,
          Calcutta-700009, p. 10.

\bibitem[Campbell 1926]{Campbell1926} Campbell, L. 1926. {\it Monthly Reports and Annual Report of the AAVSO}, 
[Reprinted from Popular Astronomy, Vol. XXXIV, Number 1-10].

\bibitem[Bandyopadhyay \& Chakrabarty 1991]{Bandyopadhyay1991} Bandyopadhyay, A. and  Chakrabarty, R. 1991. {\it Ind. J. Hist.
           Sc.}, {\bf 26}, p. 103.

\bibitem[Campbell 1928]{Campbell1928} Campbell, L. 1928. {\it Monthly Report of the AAVSO}, p. 9.

\bibitem[Hoffleit 1942]{Hoffleit1942} Hoffleit, D. 1942. {\it Sky and Telescope}, September, p. 6.

\bibitem[Campbell 1932b]{Campbell1932b} Campbell, L. 1932b. {\it Popular Astronomy},  {\bf 40}, p. 629.

\bibitem[Campbell 1946]{Campbell1946} Champbell, L. 1946. {\it Popular Astronomy}, {\bf 54}, p. 257.

\bibitem[Hoffleit 1946]{Hoffleit1946} Hoffleit, D. 1946. {\it Sky and Telescope}, June, p. 3. 

\bibitem[Eaton 1922]{Eaton1922} Eaton, H.O. 1922. {\it Monthly Report of the AAVSO}, [Reprinted from Popular Astronomy, {\bf XXX} \&
            {\bf XXXI}].

\bibitem[Struve et al. 1962]{Struve1962} Struve, O. et al. 1962. {\it Astronomy of the 20th Century}, Macmillan Co., N.Y., p. 354.

\bibitem[Mayall 1963]{Mayall1963} Mayall M.W. 1963. {\it J. R. Astron. Soc. Canada},  {\bf 57}, p. 279.

\bibitem[Chandra 1985]{Chandra1985} Chandra, R. G. 1985. {\it Dhumketu}, Puthipatra Pvt. Ldt., Cacutta-700 009.

\bibitem[Newton 1865]{Newton1865} Newton, H.A. 1865. {\it American journal of Science}, {\bf 39}, Page-193.

\bibitem[Oliver 1931]{Oliver1931} Oliver, C. 1931. {\it The Flower Astronomical Observatory}, Reprint No. 12, P. 17
            [Reprinted from Popular Astronomy, {\bf XXXIX}].

\bibitem[van der Bilt 1934]{Bilt1934} van der Bilt, J. 1934. {\it Mon. Not. R. Astron. Soc.}, {\bf 94}, p.846.

\bibitem[Campbell 1936]{Campbell1936} Campbell, L. 1936. {\it Popular Astronomy},  {\bf 44}, p. 272.

\bibitem[Mayall 1951]{Mayall1951} Mayall M.W. 1951. {\it  Popular Astronomy}, {\bf 59}, p. 332.

\bibitem[Campbell 1941]{Campbell1941} Campbell, L. 1941. {\it Variable Star Notes from the AAVSO}, p. 50-52 [Reprinted from Popular Astronomy, {\bf XLIX}].

\bibitem[Hoffleit 1947]{Hoffleit1947} Hoffleit, D. 1947. {\it Sky and Telescope}, December, p. 7.

\bibitem[Williams \& Saladyga 2011]{Williams2011} Williams, T.R. and Saladyga, M. 2011. 
{\it Advancing Variable Star Astronomy: The Centenial History of the American Association 
of Variable Star Observers}, Cambrige Univ. Press, p. 72 - 74.

\bibitem[Chakrabarty 1994]{Chakrabarty1994} Chakrabarty, R. 1994. {\it Letter to the editor}, AAVSO, {\bf 23}.


\end{thebibliography}
\end{document}